\documentclass[aps,10pt,prd,twocolumn,superscriptaddress,preprintnumbers, showkeys,nofootinbib,showpacs]{revtex4-2}
\usepackage{graphicx}  
\usepackage{slashed}
\usepackage{eucal} 
\usepackage{bm}
\usepackage[utf8]{inputenc}
\usepackage[colorlinks=true,pdfstartview=FitV,bookmarks=true,bookmarksnumbered=true,
breaklinks]{hyperref}

\hypersetup{linkcolor=blue, citecolor=blue, urlcolor=blue}
\usepackage{amsmath,amssymb,amsfonts,latexsym}

\allowdisplaybreaks
\begin{document}
\preprint{INHA-NTG-07/2025}
\title{Quadrupole forces between quark/gluon subsystems inside
   higher-spin particles}  

\author{June-Young Kim}
\email[E-mail: ]{jykim@jlab.org}
\affiliation{Theory Center, Jefferson Lab, Newport News, VA 23606,
  USA} 
\affiliation{Department of Physics, Inha University, Incheon 22212,
  Republic of Korea}
\affiliation{CPHT, CNRS, \'Ecole polytechnique, Institut Polytechnique
  de Paris, 91120 Palaiseau, France} 

\author{Hyun-Chul~Kim}
\email[E-mail: ]{hchkim@inha.ac.kr}
\affiliation{Department of Physics, Inha University, Incheon 22212,
  Republic of Korea}
\affiliation{Physics Research Institute, Inha University, Incheon
  22212, Republic of Korea}
\affiliation{School of Physics, Korea Institute for Advanced Study 
  (KIAS), Seoul 02455, Republic of Korea}

\date{\today}
\begin{abstract}
We generalize the mechanical interpretation of the forces between
quark and gluon subsystems, previously studied for the nucleon, to
arbitrary higher-spin particles. For spin-0 and spin-1/2 particles,
this force is characterized by the non-conserved $\bar{c}(t)$ form
factor. However, such an interpretation has not yet been established
for higher-spin particles due to the intricate structure of the
non-conserved energy-momentum tensor (EMT) form factors. By performing
a multipole expansion, we identify the physically meaningful
combinations of the non-conserved covariant EMT 
form factors and provide them with a clear mechanical interpretation.  
\end{abstract}
\pacs{}
\keywords{}
\maketitle
\tableofcontents
\section{Introduction}
The energy momentum tensor (EMT) has long been regarded as a valuable 
tool for investigating the quark and gluon structure of hadrons. In
the 1960s~\cite{Kobzarev:1962wt, Pagels:1966zza}, it was considered
largely an academic subject, despite containing rich physical
information. With the realization that the EMT can be accessed through
hard exclusive scattering processes, however, it has since become an
important focus of research. Its interpretation in terms of energy,
angular momentum, and mechanical distributions has been extensively
developed~\cite{Polyakov:2002yz, Goeke:2007fp, Polyakov:2018zvc,
  Lorce:2018egm, Panteleeva:2021iip, Kim:2021jjf, Freese:2021czn,
  Burkert:2023wzr, Lorce:2025oot}. The EMT structure has also been
studied for higher-spin particles~\cite{Polyakov:2019lbq, Kim:2020lrs,
  Kim:2022wkc, Freese:2022yur, Dehghan:2023ytx} and in
transitions~\cite{Kim:2022bwn, Kim:2023xvw, Kim:2023yhp,
  Alharazin:2023zzc, Kim:2025ilc, Kim:2025ves}, opening a novel avenue
for accessing tensor-polarized (quadrupole) structures in hadrons with
$J \geq 1$. 

Unlike the electromagnetic structure, the EMT provides separate
information on quark and gluon contributions, thereby emphasizing the
role of gluons and motivating the physics program of the Electron Ion
Collider (EIC). In this context, understanding the interplay between
quarks and gluons is essential, as it is characterized by the
non-conserved EMT form factors~\cite{Polyakov:2018exb}. These form
factors have been interpreted in terms of the forces between quark and
gluon subsystems~\cite{Polyakov:2018exb}, and this concept was
subsequently extended to the forces between different quark
flavors~\cite{Won:2023cyd}. Subsequent studies for spin-1/2 particles
have also been reported~\cite{Won:2023zmf, Freese:2024rkr,
  Ji:2025gsq}. 

For spin-0 and spin-1/2 particles, information on these forces is
encoded in a single $\bar{c}(t)$ form factor, making their
interpretation straightforward and simple. However, for higher-spin
particles, a corresponding interpretation of the non-conserved EMT
form factors is still lacking. Moreover, the structure of higher-spin
particles is far more intricate, requiring a systematic organization
with clear physical meaning. In this work, we aim to complete this
mechanical interpretation for arbitrary-spin particles by employing a
multipole expansion and establishing its connection to the
non-conserved EMT form factors. 

The current work is organized as follows. In Section~\ref{sec:2}, we
introduce the definition of the EMT operator in QCD and interpret its
hadronic matrix element in terms of three-dimensional (3D) EMT
distributions. The method of multipole expansion is outlined in
Section~\ref{sec:3} and applied to the stress tensor in
Section~\ref{sec:4}. In Section~\ref{sec:5}, we extend the mechanical
interpretation of the forces between the quark and gluon subsystems to
higher-spin particles within the framework of the multipole
expansion. Their connection to the non-conserved EMT form factors is
established for particles with $J < 2$ in Section~\ref{sec:6}. Using a
toy model, we then visualize the forces for a spin-1 particle in both
longitudinally and transversely polarized states in
Section~\ref{sec:7}. Finally, we present our conclusions and discuss
future outlook.

\section{Energy-momentum tensor \label{sec:2}}
The quark~($Q$) and gluon~($g$) parts of the EMT operator in QCD are
given by~\cite{Polyakov:2018zvc} 
\begin{subequations}
\label{eq:QCD_op}
\begin{align}
\hat{T}^{\mu \nu}_{Q}(x) &= \frac{1}{4} \bar{\psi}_{Q}(x) \gamma^{
 \{\mu} i \overleftrightarrow{\nabla}^{\nu\}
 } \psi_{Q}(x),  \\ 
\hat{T}^{\mu \nu}_{g}(x) &= - F^{c\mu\lambda}(x) F^{c \nu}_{\ \;
  \lambda}(x) + \frac{1}{4}g^{\mu \nu}
  F^{2}(x), 
\end{align}
\end{subequations}
where $a^{\{\mu} b^{\nu\}} = a^{\mu} b^{\nu} + a^{\nu} b^{\mu}$, and
$\overleftrightarrow{\nabla} = \overrightarrow{\nabla} -
\overleftarrow{\nabla}$ denotes the covariant derivative. Here, $F$
represents the field strength tensor. 
Note that the EMT current is conserved only when both quark
and gluon contributions are combined: 
\begin{align}
\label{eq:CC}
\partial_{\mu}\hat{T}^{\mu \nu}(x) &= 0, \\[1.5ex]
\mathrm{with} \quad \hat{T}^{\mu \nu}(x)  &= \sum_{a=Q,g} \hat{T}^{\mu
  \nu}_{a}(x). \nonumber 
\end{align}
The matrix element of the QCD EMT operator~\eqref{eq:QCD_op} between
any hadron states encodes information about the internal structure and
is parameterized by the EMT form factors. It is defined as 
\begin{align}
\label{eq:ME}
\langle p', s' | \hat{T}^{\mu \nu} (0) | p ,s  \rangle,
\end{align}
where the initial $| p, s \rangle$ and final $\langle p', s'|$ hadron
states are characterized by their momenta $p$ and $p'$ and spin
polarizations $s$ and $s'$, respectively. The states are
normalized as $\langle p', s'| p, s \rangle = 2E \,
(2\pi)^{3}\delta^{(3)}({\bm{p}'-\bm{p}}) \, \delta_{s's}$ where $E$ is
the average energy of the initial and final states.

The momentum variables can be expressed in terms of the average and
difference of the initial and final momenta as 
\begin{align}
P = \frac{p'+p}{2}, \quad \Delta = p'-p.
\label{eq:variable}
\end{align}
The on-shell condition for the initial and final states, $p'^2 = p^2 =
m^2$, can then be rewritten in terms of the variables in
Eq.~\eqref{eq:variable} as 
\begin{align}
P \cdot \Delta =0, \quad   P^{2} + \frac{\Delta^{2}}{4} = m^{2},
\label{eq:os}
\end{align}
where $m$ denotes the hadron mass.

The matrix element of the EMT operator between hadronic
states~\eqref{eq:ME} conveys information on both the motion of a
hadron and its internal structure. By performing a
Fourier transform, one obtains the corresponding distribution in
position space, which reveals the mechanical properties inside the
hadron. However, to achieve a strict definition~\cite{Freese:2021czn}
of this spatial distribution that reflects only the intrinsic
structure, one must consider relativistic two-dimensional~(2D)
densities defined within light-front quantization and in the
corresponding frames. These densities are defined in the transverse
plane, and thus one may lose information about the longitudinal
direction. 

On the other hand, the standard frame for studying hadronic matrix
elements and their spatial distributions is the three-dimensional (3D)
Breit frame (BF)~\cite{Polyakov:2002yz}. The spatial
distribution defined in this frame is contaminated by relativistic
motion of the hadron so that it does not represent purely internal
structure and is often regarded as an ambiguous one. Nevertheless, 
the 3D interpretation provides intuitive and transparent understanding
of the corresponding internal structure. It also retains 
information along the longitudinal direction. In particular, it is
conceptually useful to interpret this 3D distribution as an
unambiguous relativistic density in the Wigner
sense~\cite{Lorce:2020onh, Won:2025dgc}. Moreover, it becomes 
exact in the large-$N_c$ limit~\cite{Lorce:2022cle} or for
non-relativistic hadronic motion. 

The connection between the 3D distributions and 2D densities has been
explored in terms of the Abel transformation~\cite{Panteleeva:2021iip,
  Kim:2021kum, Kim:2021jjf}. In particular, the physical
interpretation of the stress part of the EMT for the nucleon remains
intact, with the difference being purely geometric and only minimally
affected by the Wigner rotation. Of course, for higher-spin particles
this feature becomes more intricate~\cite{Kim:2022bia}, but in
principle the 2D densities can be reconstructed from the 3D
distributions in terms of the Abel transformation. Therefore, we will
analyze the matrix element in the 3D BF in the current work. 

The 3D BF is defined through the following symmetric frame:
\begin{align}
\label{eq:BF}
P = (E, \bm{0}), \quad \Delta = (0, \bm{\Delta}), 
\end{align}
where the on-shell condition~\eqref{eq:os} gives $E = \sqrt{m^{2} -
  t/4}$ with $\Delta^{2} = t$. In the Wigner
sense~\cite{Lorce:2020onh}, the 3D EMT distributions are defined as 
\begin{align}
\label{eq:3Ddis}
&T^{\mu \nu}_{a} (\bm{r}, s', s) \nonumber \\[0.5ex] 
&=\int \frac{d^{3}\Delta}{(2\pi)^{3} 2E}
  e^{- i \bm{\Delta} \cdot \bm{r}} \langle p', s' | \hat{T}^{\mu
  \nu}_{a} (0) | p ,s  \rangle. 
\end{align}
In particular, covariant current conservation~\eqref{eq:CC} ensures
the conservation of momentum~($T^{i0}$) and the stress
tensor~($T^{ij}$) in the BF: 
\begin{align}
\label{eq:cc_bf}
\partial_{i} T^{i0}(\bm{r}, s', s) = 0, \quad \partial_{i}
  T^{ij}(\bm{r}, s', s) = 0. 
\end{align}

\section{Multipole expansion \label{sec:3}}
The basic idea of the multipole expansion is as follows: the 3D
angular momentum of the operator~\eqref{eq:3Ddis} is constructed from
a rank-$n$ irreducible tensor of the 3D position vector $\bm{r}$ and
the polarization operator $T_{LM}$~\cite{Varshalovich:1988ifq}, which
conveys information on the spin-polarization dependence $s'$ and
$s$. In addition, the parity of the structure is controlled by the
antisymmetric tensor 
$\epsilon^{ijk}$. Time-reversal symmetry and Hermiticity further
constrain the allowed structures. For a covariant multipole expansion,
see Ref.~\cite{Cotogno:2019vjb}. 

To carry this out in the basis of 3D angular momentum, we first
decompose the covariant $\mu\nu$-components of the 3D EMT
distribution~\eqref{eq:3Ddis} into $00$-, $0i$-, and $ij$-components 
according to their angular momenta: 
\begin{subequations}
\label{eq:op}
\begin{align}
&T^{00}  && (L=0), \\[1ex]
&T^{0i}  && (L=1), \\[1ex]
&T^{ij}  && (L=0,2).
\end{align}
\end{subequations}
These components are interpreted as the distributions of energy,
angular momentum, and stress tensor, respectively.  

Next, depending on the spin quantum number of the hadron considered,
the matrix element of the polarization operator is determined by 
\begin{align}
(T_{L M})_{s's} \propto C^{J s'}_{J s L M},
\label{eq:polar}
\end{align}
where the maximal rank of the polarization operator matrix elements is
dictated by the triangular selection rule of the Clebsch-Gordan (CG)
coefficient $C^{J s'}_{J s L M}$. For example, the maximally allowed
multipole spin structures for spin-1 and spin-3/2 particles correspond
to quadrupole ($L=2$) and octupole ($L=3$) spin operators,
respectively. For an arbitrary hadron with spin $J$, the multipole
spin structure extends up to $L \leq 2J$ as follows: 
\begin{subequations}
\label{eq:b}
\begin{align}
\delta_{s's} &\equiv \bm{1} &&(L=0), \\[0.5ex]
\hat{J}^{i}_{s's} &\equiv J^{i} &&(L=1), \\[0.5ex]
\hat{Q}^{ij}_{s's} &\equiv Q^{ij} &&(L=2), \\ 
\vdots \nonumber \\
\hat{M}^{i_{1}...i_{L}}_{s's}&\equiv M^{i_{1}...i_{L}} &&(L=2J), 
\end{align}
\end{subequations}
In this work, we explicitly present terms up to the quadrupole
structure. The matrix element of the dipole spin operator ($L=1$) is
defined in the spherical basis as 
\begin{align}
\label{eq:spin}
\hat{J}^{M}_{s's}= \sqrt{J(J+1)} C^{J s'}_{J s 1 M} \quad (M=0,\pm1).
\end{align}
The quadrupole spin operator can be defined in terms of the spin
operator~\eqref{eq:spin} as 
\begin{align}
&\hat{Q}^{ij}_{s's} = \frac{1}{2} \left( \hat{J}^{i} \hat{J}^{j} +
  \hat{J}^{j} \hat{J}^{i} - \frac{2}{3} J(J+1) \right)_{s's}, 
\end{align}
with $i,j=1,2,3$, which is manifestly traceless and symmetric:
\begin{align}
\hat{Q}^{ij}_{s's} = \hat{Q}^{ji}_{s's}, \quad \hat{Q}^{ii}_{s's}=0.
\end{align}

According to the allowed maximal rank of the spin
operator~\eqref{eq:b} and the angular momentum of the operator in
Eq.~\eqref{eq:op}, the maximal rank of the irreducible tensor of the
3D position vector $\bm{r}$ (orbital angular momentum) is determined.
This is referred to as the angular momentum selection rule. The
rank-$n$ irreducible tensors for the position vector are defined as 
\begin{align}
Y^{i_1 ... i_n}_{n} = \frac{(-1)^{n}}{(2n+1)!!} r^{n+1} \partial^{i_1}
  ... \partial^{i_n} \frac{1}{r}, 
\end{align}
with $r = |\bm{r}|$. More explicit expressions are given by 
\begin{subequations}
\label{eq:a}
\begin{align}
&Y_{0} = 1, && (L=0), \\[1.8ex]
&Y^{i}_{1} = \hat{r}^{i} && (L=1), \\[0.5ex]
&Y^{ij}_{2} = \hat{r}^{i}\hat{r}^{j} - \frac{1}{3}\delta^{ij} &&
     (L=2), \\[0.8ex]
&Y^{ijk}_{3} = \hat{r}^{i}\hat{r}^{j}\hat{r}^{k}  \cr
&\hspace{0.7cm}- \frac{1}{5}
  (\delta^{ij}\hat{r}^{k} + \delta^{ik}\hat{r}^{j} +
  \delta^{jk}\hat{r}^{i}) && (L=3), 
\end{align}
\end{subequations}
with $\bm{\hat{r}} = \bm{r}/|\bm{r}|$. By employing the angular
momentum selection rules and discrete symmetries~(parity, hermiticity,
time-reversal), one can perform the multipole expansion of
Eq.~\eqref{eq:op} in the basis of Eqs.~\eqref{eq:b} and \eqref{eq:a}. 

\section{Multipole expansion for stress tensor \label{sec:4}}
In this work, we focus on the stress tensor component of the EMT for a
particle with positive intrinsic parity. After performing the
multipole expansion, the stress tensor for an arbitrary-spin particle
can be expressed in the multipole basis~\cite{Panteleeva:2020ejw} as 
\begin{align}
T^{ij}_{a}(\bm{r},s',s)  &= \bm{1} \, \delta^{ij} {p}^{a}_{0}(r) +
                           \bm{1} \, Y^{ij}_{2} 
  {s}^{a}_{0}(r) \nonumber  \\[0.5ex]
&+ Q^{ij} \bigg{(}{p}^{a}_{2}(r) +
  \frac{1}{3}p^{a}_{3}(r) - \frac{1}{9}s^{a}_{3}(r) \bigg{)} \cr 
&+ 2\bigg{[} Q^{ip}Y_{2}^{pj} + Q^{jp}Y_{2}^{pi}
  -\delta^{ij}Q^{pq}Y_{2}^{pq}\bigg{]} \cr 
&\times
  \bigg{(}{s}^{a}_{2}(r)-\frac{1}{2}p^{a}_{3}(r)+\frac{1}{6}s^{a}_{3}(r)\bigg{)}
 \cr
  &+ Q^{pq}Y^{pq}_{2} \bigg{[} \left( \frac23 p^{a}_{3}(r) + \frac19
  s^{a}_{3}(r)\right) \delta^{ij} \nonumber  \\[0.3ex]
  & + \left( \frac12 p^{a}_{3} (r) +
  \frac56 s^{a}_{3}(r) \right) Y^{ij}_{2}\bigg{]} +\cdots ,
\label{eq:m_den}
\end{align}
where the ellipsis denotes contributions from spin operators
beyond the quadrupole, such as the hexadecapole (16-pole), which
become relevant for particles with $J \geq 2$.  
As a consequence of discrete symmetries, only even-rank multipole spin
operators and orbital angular momentum structures contribute to
$T^{ij}$ for spin-1 and spin-3/2 particles.
The triangular condition~\eqref{eq:polar}, together with the angular
momentum selection rule, implies that the maximal ranks of the spin
operator and orbital angular momentum are $L \leq 2$ and $L \leq 4$,
respectively.  

The dynamical information is carried by the sets of pressure
$p_{n}(r)$ and shear-force $s_{n}(r)$ distributions. The functions
$p_{0}(r)$ and $s_{0}(r)$ describe the pressure and shear-force
distributions in spherically symmetric hadrons. Since higher-order
sets appear as additional structures for higher-spin particles, they
are referred to as the quadrupole (or higher-multipole) pressure and
shear-force distributions~\cite{Panteleeva:2020ejw}.  

These higher-order sets are constructed under the constraint of current
conservation~\eqref{eq:cc_bf} and naturally preserve the relation
between pressure and shear forces that holds for spherically symmetric
hadrons (spin-0 and spin-1/2 particles). For example, by
taking derivatives of the stress tensor~\eqref{eq:m_den}, one finds
that the sets of quadrupole pressure and shear-force distributions
also satisfy the following equilibrium
equations~\cite{Panteleeva:2020ejw}, which are likewise valid in the 
monopole sector: 
\begin{align}
\label{eq:EMT_differential}
&\sum_{a=Q,g} \left[\frac{2}{3}\frac{d s^{a}_{n}(r)}{dr} + 2
  \frac{s^{a}_{n}(r)}{r} +\frac{d p^{a}_{n}(r)}{dr}\right] = 0,
  \\[1ex] 
&\mathrm{with} \quad n=0,2,3. \nonumber
\end{align}
These equations allow us to define the pressure and shear-force
distributions, $p_{n}(r)$ and $s_{n}(r)$, as the Fourier transforms of
the generalized $D$-term form factors~\cite{Panteleeva:2020ejw}:
\begin{subequations}
\label{eq:ps_FT}
\begin{align}
p_{n}(r) &= \sum_{a=Q,g}p^{a}_{n}(r) 
= \frac{1}{6m} \,\partial^{2}\mathcal{D}_{n}(r), \\ 
s_{n}(r) &= \sum_{a=Q,g}s^{a}_{n}(r) 
= -\frac{1}{4m} \, r \frac{d}{dr} \frac{1}{r} \frac{d}{dr}
  \mathcal{D}_{n}(r),
\end{align}
\end{subequations}
which immediately imply the von Laue condition for the pressure:
\begin{align}
\int d^{3}r \, p_{n}(r) 
= \frac{1}{6m} \int d^{3}r \, \partial^{2}\mathcal{D}_{n}(r) = 0.
\end{align}
Similarly, the total $r^{2}$-weighted pressure and shear-force
distributions are related to the generalized
$D$-term~\cite{Panteleeva:2020ejw}:
\begin{align}
D_{n} &\equiv \int d^{3}r \, \mathcal{D}_{n}(r) \cr
&= m \int d^{3}r \, r^{2} p_{n}(r)
= -\frac{4}{15} m \int d^{3}r \, r^{2} s_{n}(r).
\end{align} 

\section{Forces between the quark/gluon subsystems \label{sec:5}}
Once the quark and gluon contributions to the pressure and shear-force
distributions are separated, the equilibrium
equations~\eqref{eq:EMT_differential} are no longer satisfied for
quarks and gluons individually. A balanced system emerges only when
both contributions are combined. From the viewpoint of the quark
subsystem, the gluon contribution may be regarded as an external
force. This external force field $-\bm{f}$ can be defined
as~\cite{landau1986theory, Polyakov:2018exb} 
\begin{align}
\partial_{i} T^{ij}_{Q}(\bm{r},s',s) = f^{j}(\bm{r},s',s).
\label{eq:ge}
\end{align}
In classical continuum mechanics, this corresponds to the Cauchy
momentum equation for the static stress tensor. For higher-spin
particles, however, quantum effects become evident in the
spin-polarization dependence of the distributions~\eqref{eq:ge},
which may be considered as a natural generalization to quantum
systems.  

This external force field can be expanded in terms of the multipole
structures using the bases of Eqs.~\eqref{eq:b} and \eqref{eq:a}. More
explicitly, by applying derivatives to the stress
tensor~\eqref{eq:m_den}, following multipole structures appear: 
\begin{align}
\label{eq:mul_for}
f^{j}(\bm{r},s',s) &= \bm{1} \, Y^{j}_1 f_{0}(r) + Q^{jk} Y^{k}_1
  f_{2}(r) \nonumber \\[1ex]
  &+ Q^{kl} Y^{klj}_3 f_{3}(r) + \cdots, 
\end{align}
which are induced from Eq.~\eqref{eq:m_den}. Consequently, the number
of multipole structures in Eq.~\eqref{eq:mul_for} must coincide with
the number of independent equilibrium equations in
Eq.~\eqref{eq:EMT_differential}. The function $f_{0}(r)$ originates
from the monopole sector of the pressure and shear-force distributions
and can therefore be regarded as the monopole force distribution. In
contrast, $f_{2}(r)$ and $f_{3}(r)$ stem from the quadrupole sector of
the pressure and shear-force distributions and are thus identified as
quadrupole force distributions. Similar to the construction of the
pressure and shear-force distributions~\eqref{eq:ps_FT}, the multipole
distributions for Eq.~\eqref{eq:mul_for} can be derived as
\begin{subequations} 
\label{eq:F_dis}
\begin{align}
f_{0}(r) &= - m \frac{d}{dr} \bar{\mathcal{C}}_{0}(r), \\[0.5ex]
f_{2}(r) &= - m \frac{d}{dr} \bar{\mathcal{C}}_{2}(r), \\
f_{3}(r) &=  \frac{1}{m} r^{3}
           \left(\frac{1}{r}\frac{d}{dr}\right)^{3}
           \bar{\mathcal{C}}_{3}(r),  
\end{align}
\end{subequations}
where $\bar{\mathcal{C}}_{n}(r)$ denotes the Fourier transform of the
non-conserved multipole EMT form factors $\bar{C}^{Q}_{n}(t)$
generalized $\bar{c}(t)$ form factor) parameterizing the hadronic
matrix element~\eqref{eq:ME}: 
\begin{align}
\bar{\mathcal{C}}_{n}(r) =  \int \frac{d^{3}\Delta}{(2\pi)^{3}}
  e^{-i\bm{\Delta}\cdot \bm{r}} \bar{C}^{Q}_{n}(t) \quad (n=0,2,3). 
\end{align}
The conservation of the total EMT~\eqref{eq:cc_bf}
  implies that the sum of the quark and gluon contributions to the
  non-conserved multipole EMT form factors vanishes, leading to 
\begin{align}
\label{eq:glu_cbar}
\bar{C}^{Q}_{n}(t) =-\bar{C}^{g}_{n}(t).
\end{align}
More importantly, the definitions of the separate quark and gluon
parts of the non-conserved form factors depend on the chosen
decomposition of the QCD EMT operator~\cite{Ji:1995sv, Liu:2021gco,
  Lorce:2017xzd, Hatta:2018sqd, Ji:2025qax}; in particular, these
choices are renormalization-scheme dependent~\cite{Hatta:2018sqd,
  Metz:2020vxd}. To avoid this issue, one may decompose the EMT into
quark, gluon, and trace-anomaly parts~\cite{Ji:1995sv,
  Liu:2021gco}. Among these decompositions, the definition of the
quark $\bar{C}^{Q}$ form factor is less ambiguous, so that we will
adopt this part in what follows. The superscript $Q$ on the
non-conserved form factors will be omitted for brevity, unless 
explicitly required.  

From Eq.~\eqref{eq:F_dis}, the multipole moments of the force
distributions can be related to the $\bar{C}_{n}(0)$ form factors as  
\begin{subequations}
\begin{align}
\int d^{3}r \; r \, f_{0,2}(r) &= 3 m \, \bar{C}_{0,2}(0), \\
\int d^{3}r \; r^{3} \, f_{3}(r) &= - \frac{105}{m} \, \bar{C}_{3}(0).
\end{align}
\end{subequations}

The ellipses in Eq.~\eqref{eq:mul_for} correspond to higher-multipole
structures relevant for particles with even larger spin ($J \geq
2$). Incorporating Eq.~\eqref{eq:F_dis}, we express these higher-order
multipole distributions generically as 
\begin{align}
f(r)=\frac{i^{N+1}}{m^{N-2}} r^{N} \left(\frac{1}{r}
  \frac{d}{dr}\right)^{N} \bar{\mathcal{C}}(r)  \quad
  (N=\mathrm{odd}), 
\end{align}
where $f(r)$ designates the multipole distribution associated with a
rank-$N$ irreducible tensor in the expansion~\eqref{eq:mul_for}. The
corresponding multipole moment is then given by 
\begin{align}
&\int d^{3}r \; r^{N} \, f(r) \nonumber \\[0.5ex]
&= - \frac{(2N+1)!!}{m^{N-2}} i^{N+1} \,
  \bar{C}(0)   \quad (N=\mathrm{odd}). 
\end{align}

By projecting $f^{j}$ in Eq.~\eqref{eq:mul_for} onto spherical
components, the force can be decomposed into normal and tangential
parts: 
\begin{align}
\label{eq:force_sp}
\bm{f}(\bm{r},s',s) &= f^{r}(\bm{r},s',s) \, \bm{\hat{r}} \nonumber \\[1ex]
&+
  f^{\theta}(\bm{r},s',s) \, \bm{\hat{\theta}}  + f^{\phi}(\bm{r},s',s)
  \, \bm{\hat{\phi}}. 
\end{align}
The explicit forms of the normal and tangential distributions are
\begin{subequations}
\label{eq:force_sp_2}
\begin{align}
f^{r}(\bm{r},s',s) &= \bm{1} \, f_{0}(r) \nonumber \\[0.5ex]
&+ Q^{rr} \, \left[f_{2}(r) +
                     \frac{3}{5}f_{3}(r) \right] + \cdots, \\[0.5ex] 
f^{\theta}(\bm{r},s',s) &=  Q^{r\theta} \, f_{2}(r) + \cdots, \\[1.5ex] 
f^{\phi}(\bm{r},s',s) &=  Q^{r\phi} \, f_{2}(r) + \cdots,  
\end{align}
\end{subequations}
with $Q^{r\theta} \equiv Q^{ij} \hat{r}^{i} \, \hat{\theta}^{i}$.
The total force is then defined by integrating Eq.~\eqref{eq:force_sp}
contracted with the spherical vectors $\bm{\hat{v}}$ over 3D
space~\cite{Polyakov:2018exb}: 
\begin{align}
\label{eq:total_f}
F^{v}(s',s)&= \int d^{3}r \, \bm{\hat{v}} \cdot \bm{f}(\bm{r},s',s)
             \quad (v=r, \theta, \phi). 
\end{align}
For $v = \theta,\phi$, the angular integration over $\theta$ and
$\phi$ causes the averaged force to vanish. For $v = r$, the angular
integration eliminates contributions from higher-multipole structures,
e.g, $Q^{rr}$, so that the monopole component alone determines the
averaged force~\eqref{eq:total_f}. 

In classical continuum mechanics, or for spherically symmetric
hadrons, the external force manifests solely as a normal
component. By contrast, for higher-spin particles ($J>1/2$),
tangential components of the force also appear due to spin-polarization 
dependence, reflecting the more intricate interplay between the quark
and gluon quantum subsystems. 

\section{Connection to the covariant form factors \label{sec:6}}
The dynamical information of the multipole force distributions
$f_{0,2,3}$ is carried by the non-conserved EMT form factors. In
several works~\cite{Polyakov:2018zvc, Polyakov:2019lbq, Kim:2020lrs,
  Cotogno:2019vjb, Tanaka:2018wea, Cosyn:2019aio}, these form factors
are defined through the covariant parameterization of the hadronic
matrix element. In this section, we relate these covariant form
factors to our multipole form factors $\bar{C}_{n}(t)$ for particles
with $J<2$. Making use of the relation between the force distribution
and the hadronic matrix element, we establish the following relations: 
\begin{align}
&f^{j}(\bm{r},s',s) \nonumber \\[1ex]
&= - i  \int \frac{d^{3}\Delta}{(2\pi)^{3}2E} e^{-i
  \bm{\Delta} \cdot \bm{r}} \Delta^{i} \langle p',s' |
  \hat{T}^{ij}_{Q}(0) | p,s \rangle.
\end{align}

\textit{Spin-0 particle.} Due to the triangular condition in
Eq.~\eqref{eq:polar}, the only allowed multipole spin operator is $L=0$,
which implies that all quadrupole contributions vanish. The monopole form
factor $\bar{C}_{0}(t)$ can then be expressed in terms of the covariant
form factors, following the notation of Refs.~\cite{Polyakov:2018zvc,
  Tanaka:2018wea}: 
\begin{subequations}
\begin{align}
\bar{C}_{0}(t) &=  \frac{m}{2E}  \bar{c}(t), \\[0.5ex]
\bar{C}_{2}(t)&=\bar{C}_{3}(t) = 0.
\end{align}
\end{subequations}

\textit{Spin-1/2 particle.} Similar to the spin-0 case, only even-rank
multipole spin operators are permitted by discrete
symmetries. Therefore, in the stress tensor, only the monopole spin
operator contributes, while the quadrupole structure is 
absent. Using the parameterization of Ref.~\cite{Polyakov:2018zvc},
the non-conserved monopole form factor $\bar{C}_{0}(0)$ is related
to the covariant form factor as follows: 
\begin{subequations}
\label{eq:spin12}
\begin{align}
\bar{C}_{0}(t) &=  \bar{c}(t), \\[1.5ex]
\bar{C}_{2}(t)&=\bar{C}_{3}(t) = 0.
\end{align}
\end{subequations}

\textit{Spin-1 particle.} The quadrupole spin operator first arises in
the case of spin-1 particles, offering a novel perspective on 
quadrupole structures in the force distribution. In the notation of
the covariant form factors of Ref.~\cite{Polyakov:2019lbq}, the
multipole form factors $\bar{C}_{n}(t)$ are related to the covariant
form factors as follows: 
\begin{subequations}
\begin{align}
\bar{C}_{0}(t) &= \frac{m}{2E} \bigg{[} \left( 1 -
                 \frac{t}{6m^{2}}\right)\bar{c}_{0}(t) \cr 
& -  \frac{t}{12m^{2}}  \left(1 - \frac{t}{4m^{2}}\right) \bar{c}_1
  (t)  \nonumber \\
  &- \frac{1}{6}\left( 1 - \frac{t}{2m^{2}}\right)\bar{f} (t)
  \bigg{]}, \\ 
\bar{C}_{2}(t) &= \frac{m}{2E} \bigg{[}  \frac{t}{5m^{2}}\bar{c}_{0}
                 (t)  \cr
&+ \frac{t}{10m^{2}} \left(1 -
                 \frac{t}{4m^{2}}\right)\bar{c}_1 (t) \cr 
&+\bigg{\{}  \frac{20m(m+E)-7t}{10m (m+E)} \cr
&+
  \frac{t}{10m^{2}} \bigg{(}1 + \frac{t}{2(m+E)^{2}}\bigg{)}  \bigg{\}}
  \bar{f} (t) \bigg{]}, \\ 
\bar{C}_{3}(t) &= - \frac{m}{2E}\bigg{[} \frac{1}{2}\bar{c}_0 (t)  +
                 \frac{1}{4} \left( 1 - \frac{t}{4m^{2}}\right)
                 \bar{c}_1 (t) \cr 
&+ \frac{2E^{2} - 2m^{2} + t}{8 (m + E)^{2}} \bar{f} (t)     \bigg{]}.  
\end{align}
\end{subequations}

\textit{Spin-3/2 particle.}
As in the spin-1 case, only even-rank
multipole spin operators are allowed by discrete symmetries. Thus, the
stress tensor exhibits the same multipole structure as that of the
spin-1 particle. The non-conserved multipole EMT form factors can be
expressed in terms of the notation of
Refs.~\cite{Kim:2020lrs, Cotogno:2019vjb}:
\begin{subequations}
\begin{align}
\bar{C}_{0}(t) &= \left( 1 - \frac{t}{6m^{2}}\right)F_{3,0}(t)  \cr
& - \frac{t}{6m^{2}}  \left(1 - \frac{t}{4m^{2}}\right) F_{3,1}(t), \cr 
&+                \frac{2}{3}\left( 1 -
                 \frac{t}{4m^{2}}\right)F_{6,0}(t) \\
\bar{C}_{2}(t) &=  \frac{t}{15m^{2}}F_{3,0}(t) \cr
&+ \frac{t}{15m^{2}} \left( 1 -
  \frac{t}{4m^{2}} \right)F_{3,1}(t), \cr 
&-\frac{2}{3} \left(1 + \frac{t \{ t-14m(m+E) \} }{40
                 m^{2} (m+E)^{2}} \right) F_{6,0}(t), \\
\bar{C}_{3}(t) &= - \frac{1}{6}F_{3,0}(t) \cr
&- \frac{1}{6}\left(1-
                 \frac{t}{4m^{2}}\right)F_{3,1}(t) \cr 
&- \frac{1}{6}\left(\frac{m}{m+E} -
  \frac{t}{4(m+E)^2}\right)F_{6,0}(t). 
\end{align}
\end{subequations}

\textit{Spin-2 particle.} For a spin-2 particle, the covariant form
factors can be related to multipole form factors; however, this
requires going beyond the quadrupole spin operator to, for example,
the hexadecapole (rank-4) operator. This implies that orbital angular
momenta up to rank-6 may contribute to the stress tensor due to the
angular momentum selection rules. Five non-conserved form factors
appear in the parameterization of the matrix
element~\cite{Cotogno:2019vjb}, which may additionally give rise to
two more multipole form factors, $\bar{C}_{n}(t)$ ($n=4,5$), compared
to spin-1 and spin-3/2 particles. Such a structure is particularly
interesting for probing the internal mechanical properties of spin-2
particles such as tensor mesons. 

\section{Numerical results and visualization \label{sec:7}}
To visualize the force distributions characterized by the
non-conserved EMT form factors, we perform a numerical
estimation. Specifically, we consider a spin-1 particle polarized both
longitudinally and transversely. For a spin-3/2 particle, the
quadrupole pattern is the same as that of the spin-1 particle, but its
strength depends entirely on dynamical information. Since this
information is currently unknown, we focus on the spin-1 particle as a
representative example. 
\begin{figure}[t]
\centering
\includegraphics[scale=0.61]{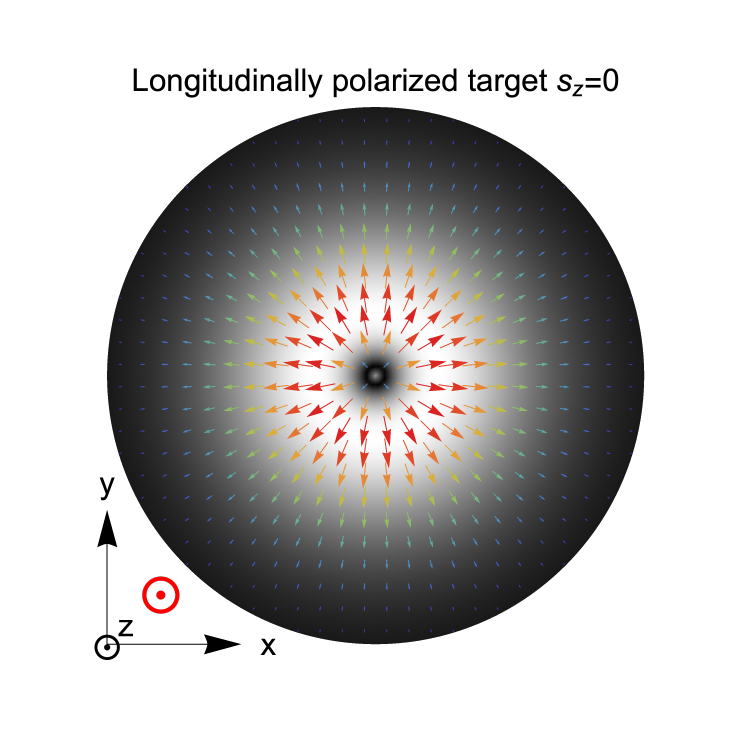}
\includegraphics[scale=0.61]{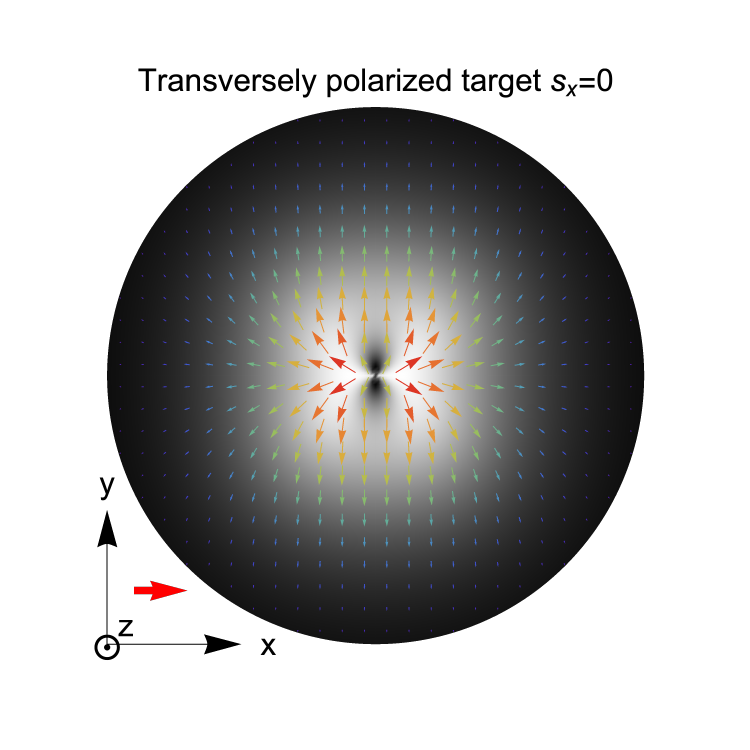}
\caption{Visualization of the force distribution for a spin-1 target
  polarized longitudinally (upper panel) and transversely (lower
  panel).} 
\label{fig:1}
\end{figure}

For the numerical analysis, we take the multipole form factors
$\bar{C}_{n}(t)$ as input and parameterize them by a tripole
form:
\begin{align}
\label{eq:astz}
\bar{C}^{Q}_n (t)= \frac{\bar{C}^{Q}_{n}(0)}{ \left(1 -
  \frac{t}{\Lambda^{2}}\right)^{3}}, 
\end{align}
with $n=0,2,3$. The mass of the spin-1 hadron and the tripole mass
are set to $m=\Lambda=1~\mathrm{GeV}$, and the normalizations of the
non-conserved form factors are taken to be   
\begin{align}
\label{eq:inst}
\bar{C}^{Q}_{0,2,3}(0) = 1.4 \times 10^{-2}.
\end{align}
The value of $\bar{C}^{Q}_{0}(0)=\bar{c}^{Q}(0)$ [see
Eq.~\eqref{eq:spin12}] for the nucleon was predicted from the QCD
instanton vacuum~\cite{Polyakov:2018exb}, and we conjecture that this
small and positive behavior persists for the higher multipole form
factors as well. As discussed in Eq.~\eqref{eq:glu_cbar},
Eq.~\eqref{eq:inst} represents the partial form factor (force
distribution) from the quark part. In the combined quark-gluon system,
the force distribution is balanced and reaches equilibrium by
considering the gluon contribution $\bar{C}^{g}_{n}$, which has the
opposite sign to the quark contribution. 

To estimate the typical magnitude of the force distribution, we
compute the averaged force $F^{r}$. After integrating the radial force
distribution $f^{r}$ over $r$, we obtain 
\begin{align}
F^{r}(s',s) = \bm{1} \, \frac{3}{4} \Lambda \,  m  \,
  \bar{C}^{Q}_{0}(0) \approx \bm{1} \times  5.3 \times 10^{-2} \,
  \frac{\mathrm{GeV}}{\mathrm{fm}}, 
\end{align}
which is proportional to both the tripole mass $\Lambda$ and the hadron
mass $m$. As discussed in Eq.~\eqref{eq:total_f}, this total force is
determined solely by the monopole distribution $f_{0}(r)$ (or
equivalently, by $\bar{C}^{Q}_{0}(0)$).  

In Fig.~\ref{fig:1}, we illustrate the cross-sectional view ($\theta=\pi/2$)
of the force distribution~\eqref{eq:force_sp} for a spin-1 particle
when its spin is polarized longitudinally (upper panel) and
transversely (lower panel). For the longitudinally polarized spin-1
particle, the quadrupole polarization (see
Ref.~\cite{Varshalovich:1988ifq} for the explicit representation) is
given by 
\begin{subequations}
\label{eq:Q_l}
\begin{align}
&Q^{rr} = -\frac{1}{6}(1+3 \cos{2\theta}), \\
&Q^{r\theta} = \cos{\theta} \sin{\theta}, \\[1ex]
&Q^{r\phi} = 0, \\[1.5ex]
&[\mathrm{for} \ s'_{z}=s_{z} = 0]. \nonumber
\end{align}
\end{subequations}
Inserting these results~\eqref{eq:Q_l} into Eq.~\eqref{eq:force_sp_2},
we derive the force distribution: 
\begin{align}
\label{eq:force_l}
&\bm{f}(r,s'_{z}=0,s_{z}=0)\bigg{|}_{\theta=\frac{\pi}{2}} \cr
&=  \left[f_{0}(r) + \frac{1}{3}\left\{f_{2}(r) + \frac{3}{5}
  f_{3}(r)\right\} \right] \bm{\hat{r}}. 
\end{align}
For the cross-sectional view of this force, all $\phi$-dependent terms
in Eq.~\eqref{eq:Q_l} vanish, implying that the quadrupole pattern
disappears in Eq.~\eqref{eq:force_l}. Nevertheless, the quadrupole
contribution is still present but merges with the monopole force
distribution. Moreover, the tangential forces proportional to
$Q^{r\theta}$ and $Q^{r\phi}$ vanish at $\theta=\pi/2$.
As a result, only the normal force appears in the upper panel of
Fig.~\ref{fig:1}. 

For the transversely polarized spin-1 particle, i.e., $|s_{x}=0\rangle
= (|s_{z}=1\rangle -
|s_{z}=-1\rangle)/\sqrt{2}$~\cite{Carlson:2009ovh}, the quadrupole
polarization is given by~\cite{Varshalovich:1988ifq} 
\begin{subequations}
\label{eq:Q_t}
\begin{align}
&Q^{rr} = -\frac{1}{6}(1+3 \cos{2\phi}), \\
&Q^{r\theta} = 0, \\[1ex]
&Q^{r\phi} = \cos{\phi} \sin{\phi}, \\[1.5ex]
&[\mathrm{for} \ s'_{x}=s_{x} = 0]. \nonumber
\end{align}
\end{subequations}
Substituting these results~\eqref{eq:Q_t} into Eq.~\eqref{eq:force_sp_2},
we derive the force distribution. In contrast to the longitudinally
polarized case, all $\theta$-dependent terms vanish, leaving only
$\phi$-dependence. This gives rise to quadrupole deformation in the
force distribution on the cross-sectional plane, and generates
non-vanishing tangential forces.
A more intricate force distribution is evident in the lower panel of
Fig.~\ref{fig:1}, reflecting that hadron shape formation arises from a
more complex interplay between the quark and gluon subsystems compared
to the spherically symmetric case of hadrons with $J<1$. 

\section{Summary and outlook \label{sec:8}}
In this work, we generalized the mechanical interpretation of the
non-conserved $\bar{c}(t)$ form factor, originally formulated for
spherically symmetric hadrons, to higher-spin particles. To provide a
clear physical interpretation, we performed a three-dimensional
multipole expansion of the stress tensor in the Breit frame under the
constraint of current conservation. Separating the quark and gluon
contributions to the stress tensor isolates the non-conserved
component, which can be interpreted as the force distribution between
the quark and gluon subsystems and is connected to the non-conserved
multipole energy--momentum tensor~(EMT) form factors
$\bar{C}_{n}(t)$. Furthermore, these form factors were related to the
non-conserved covariant EMT form factors for particles with $J<2$.
Using a toy model, we illustrated the force distribution for a spin-1
particle polarized longitudinally and transversely. We found that, for
higher-spin particles, hadron shape formation results from a more
intricate balance between the quark and gluon subsystems compared to
the spherically symmetric case of hadrons with $J<1$.

A more conclusive visualization of the force distribution in
higher-spin particles requires an explicit determination of the
non-conserved EMT form factors, which can then serve as input for the
analysis. (For example, the chiral quark-soliton model can be used to
compute the EMT form factors etc.)

Finally, for a more precise interpretation of relativistic densities,
the present formalism should be reformulated in terms of a
two-dimensional light-front multipole expansion, which can be carried
out straightforwardly within the present framework. The physical
conclusion will remain intact in the two-dimensional
transverse plane.  

\vspace{0.8cm}

\textbf{Acknowledgments}: 
The present work was supported by the Basic Science
Research Program through the National Research Foundation of Korea
funded by the Korean government (Ministry of Education, Science and
Technology, MEST), Grant-No. RS-2025-00513982.   

\bibliography{ForceQG}

\begin{thebibliography}{45}%
\makeatletter
\providecommand \@ifxundefined [1]{%
 \@ifx{#1\undefined}
}%
\providecommand \@ifnum [1]{%
 \ifnum #1\expandafter \@firstoftwo
 \else \expandafter \@secondoftwo
 \fi
}%
\providecommand \@ifx [1]{%
 \ifx #1\expandafter \@firstoftwo
 \else \expandafter \@secondoftwo
 \fi
}%
\providecommand \natexlab [1]{#1}%
\providecommand \enquote  [1]{``#1''}%
\providecommand \bibnamefont  [1]{#1}%
\providecommand \bibfnamefont [1]{#1}%
\providecommand \citenamefont [1]{#1}%
\providecommand \href@noop [0]{\@secondoftwo}%
\providecommand \href [0]{\begingroup \@sanitize@url \@href}%
\providecommand \@href[1]{\@@startlink{#1}\@@href}%
\providecommand \@@href[1]{\endgroup#1\@@endlink}%
\providecommand \@sanitize@url [0]{\catcode `\\12\catcode `\$12\catcode
  `\&12\catcode `\#12\catcode `\^12\catcode `\_12\catcode `\%12\relax}%
\providecommand \@@startlink[1]{}%
\providecommand \@@endlink[0]{}%
\providecommand \url  [0]{\begingroup\@sanitize@url \@url }%
\providecommand \@url [1]{\endgroup\@href {#1}{\urlprefix }}%
\providecommand \urlprefix  [0]{URL }%
\providecommand \Eprint [0]{\href }%
\providecommand \doibase [0]{http://dx.doi.org/}%
\providecommand \selectlanguage [0]{\@gobble}%
\providecommand \bibinfo  [0]{\@secondoftwo}%
\providecommand \bibfield  [0]{\@secondoftwo}%
\providecommand \translation [1]{[#1]}%
\providecommand \BibitemOpen [0]{}%
\providecommand \bibitemStop [0]{}%
\providecommand \bibitemNoStop [0]{.\EOS\space}%
\providecommand \EOS [0]{\spacefactor3000\relax}%
\providecommand \BibitemShut  [1]{\csname bibitem#1\endcsname}%
\let\auto@bib@innerbib\@empty
\bibitem [{\citenamefont {Kobzarev}\ and\ \citenamefont
  {Okun'}(1962)}]{Kobzarev:1962wt}%
  \BibitemOpen
  \bibfield  {author} {\bibinfo {author} {\bibfnamefont {I.~Y.}\ \bibnamefont
  {Kobzarev}}\ and\ \bibinfo {author} {\bibfnamefont {L.~B.}\ \bibnamefont
  {Okun'}},\ }\href@noop {} {\bibfield  {journal} {\bibinfo  {journal} {Zh.
  Eksp. Teor. Fiz.}\ }\textbf {\bibinfo {volume} {43}},\ \bibinfo {pages}
  {1904} (\bibinfo {year} {1962})}\BibitemShut {NoStop}%
\bibitem [{\citenamefont {Pagels}(1966)}]{Pagels:1966zza}%
  \BibitemOpen
  \bibfield  {author} {\bibinfo {author} {\bibfnamefont {H.}~\bibnamefont
  {Pagels}},\ }\href {\doibase 10.1103/PhysRev.144.1250} {\bibfield  {journal}
  {\bibinfo  {journal} {Phys. Rev.}\ }\textbf {\bibinfo {volume} {144}},\
  \bibinfo {pages} {1250} (\bibinfo {year} {1966})}\BibitemShut {NoStop}%
\bibitem [{\citenamefont {Polyakov}(2003)}]{Polyakov:2002yz}%
  \BibitemOpen
  \bibfield  {author} {\bibinfo {author} {\bibfnamefont {M.~V.}\ \bibnamefont
  {Polyakov}},\ }\href {\doibase 10.1016/S0370-2693(03)00036-4} {\bibfield
  {journal} {\bibinfo  {journal} {Phys. Lett. B}\ }\textbf {\bibinfo {volume}
  {555}},\ \bibinfo {pages} {57} (\bibinfo {year} {2003})},\ \Eprint
  {http://arxiv.org/abs/hep-ph/0210165} {arXiv:hep-ph/0210165} \BibitemShut
  {NoStop}%
\bibitem [{\citenamefont {Goeke}\ \emph {et~al.}(2007)\citenamefont {Goeke},
  \citenamefont {Grabis}, \citenamefont {Ossmann}, \citenamefont {Polyakov},
  \citenamefont {Schweitzer}, \citenamefont {Silva},\ and\ \citenamefont
  {Urbano}}]{Goeke:2007fp}%
  \BibitemOpen
  \bibfield  {author} {\bibinfo {author} {\bibfnamefont {K.}~\bibnamefont
  {Goeke}}, \bibinfo {author} {\bibfnamefont {J.}~\bibnamefont {Grabis}},
  \bibinfo {author} {\bibfnamefont {J.}~\bibnamefont {Ossmann}}, \bibinfo
  {author} {\bibfnamefont {M.~V.}\ \bibnamefont {Polyakov}}, \bibinfo {author}
  {\bibfnamefont {P.}~\bibnamefont {Schweitzer}}, \bibinfo {author}
  {\bibfnamefont {A.}~\bibnamefont {Silva}}, \ and\ \bibinfo {author}
  {\bibfnamefont {D.}~\bibnamefont {Urbano}},\ }\href {\doibase
  10.1103/PhysRevD.75.094021} {\bibfield  {journal} {\bibinfo  {journal} {Phys.
  Rev. D}\ }\textbf {\bibinfo {volume} {75}},\ \bibinfo {pages} {094021}
  (\bibinfo {year} {2007})},\ \Eprint {http://arxiv.org/abs/hep-ph/0702030}
  {arXiv:hep-ph/0702030} \BibitemShut {NoStop}%
\bibitem [{\citenamefont {Polyakov}\ and\ \citenamefont
  {Schweitzer}(2018)}]{Polyakov:2018zvc}%
  \BibitemOpen
  \bibfield  {author} {\bibinfo {author} {\bibfnamefont {M.~V.}\ \bibnamefont
  {Polyakov}}\ and\ \bibinfo {author} {\bibfnamefont {P.}~\bibnamefont
  {Schweitzer}},\ }\href {\doibase 10.1142/S0217751X18300259} {\bibfield
  {journal} {\bibinfo  {journal} {Int. J. Mod. Phys. A}\ }\textbf {\bibinfo
  {volume} {33}},\ \bibinfo {pages} {1830025} (\bibinfo {year} {2018})},\
  \Eprint {http://arxiv.org/abs/1805.06596} {arXiv:1805.06596 [hep-ph]}
  \BibitemShut {NoStop}%
\bibitem [{\citenamefont {Lorc{\'e}}\ \emph {et~al.}(2019)\citenamefont
  {Lorc{\'e}}, \citenamefont {Moutarde},\ and\ \citenamefont
  {Trawi{\'n}ski}}]{Lorce:2018egm}%
  \BibitemOpen
  \bibfield  {author} {\bibinfo {author} {\bibfnamefont {C.}~\bibnamefont
  {Lorc{\'e}}}, \bibinfo {author} {\bibfnamefont {H.}~\bibnamefont {Moutarde}},
  \ and\ \bibinfo {author} {\bibfnamefont {A.~P.}\ \bibnamefont
  {Trawi{\'n}ski}},\ }\href {\doibase 10.1140/epjc/s10052-019-6572-3}
  {\bibfield  {journal} {\bibinfo  {journal} {Eur. Phys. J. C}\ }\textbf
  {\bibinfo {volume} {79}},\ \bibinfo {pages} {89} (\bibinfo {year} {2019})},\
  \Eprint {http://arxiv.org/abs/1810.09837} {arXiv:1810.09837 [hep-ph]}
  \BibitemShut {NoStop}%
\bibitem [{\citenamefont {Panteleeva}\ and\ \citenamefont
  {Polyakov}(2021)}]{Panteleeva:2021iip}%
  \BibitemOpen
  \bibfield  {author} {\bibinfo {author} {\bibfnamefont {J.~Y.}\ \bibnamefont
  {Panteleeva}}\ and\ \bibinfo {author} {\bibfnamefont {M.~V.}\ \bibnamefont
  {Polyakov}},\ }\href {\doibase 10.1103/PhysRevD.104.014008} {\bibfield
  {journal} {\bibinfo  {journal} {Phys. Rev. D}\ }\textbf {\bibinfo {volume}
  {104}},\ \bibinfo {pages} {014008} (\bibinfo {year} {2021})},\ \Eprint
  {http://arxiv.org/abs/2102.10902} {arXiv:2102.10902 [hep-ph]} \BibitemShut
  {NoStop}%
\bibitem [{\citenamefont {Kim}\ and\ \citenamefont
  {Kim}(2021{\natexlab{a}})}]{Kim:2021jjf}%
  \BibitemOpen
  \bibfield  {author} {\bibinfo {author} {\bibfnamefont {J.-Y.}\ \bibnamefont
  {Kim}}\ and\ \bibinfo {author} {\bibfnamefont {H.-C.}\ \bibnamefont {Kim}},\
  }\href {\doibase 10.1103/PhysRevD.104.074019} {\bibfield  {journal} {\bibinfo
   {journal} {Phys. Rev. D}\ }\textbf {\bibinfo {volume} {104}},\ \bibinfo
  {pages} {074019} (\bibinfo {year} {2021}{\natexlab{a}})},\ \Eprint
  {http://arxiv.org/abs/2105.10279} {arXiv:2105.10279 [hep-ph]} \BibitemShut
  {NoStop}%
\bibitem [{\citenamefont {Freese}\ and\ \citenamefont
  {Miller}(2021)}]{Freese:2021czn}%
  \BibitemOpen
  \bibfield  {author} {\bibinfo {author} {\bibfnamefont {A.}~\bibnamefont
  {Freese}}\ and\ \bibinfo {author} {\bibfnamefont {G.~A.}\ \bibnamefont
  {Miller}},\ }\href {\doibase 10.1103/PhysRevD.103.094023} {\bibfield
  {journal} {\bibinfo  {journal} {Phys. Rev. D}\ }\textbf {\bibinfo {volume}
  {103}},\ \bibinfo {pages} {094023} (\bibinfo {year} {2021})},\ \Eprint
  {http://arxiv.org/abs/2102.01683} {arXiv:2102.01683 [hep-ph]} \BibitemShut
  {NoStop}%
\bibitem [{\citenamefont {Burkert}\ \emph {et~al.}(2023)\citenamefont
  {Burkert}, \citenamefont {Elouadrhiri}, \citenamefont {Girod}, \citenamefont
  {Lorc{\'e}}, \citenamefont {Schweitzer},\ and\ \citenamefont
  {Shanahan}}]{Burkert:2023wzr}%
  \BibitemOpen
  \bibfield  {author} {\bibinfo {author} {\bibfnamefont {V.~D.}\ \bibnamefont
  {Burkert}}, \bibinfo {author} {\bibfnamefont {L.}~\bibnamefont
  {Elouadrhiri}}, \bibinfo {author} {\bibfnamefont {F.~X.}\ \bibnamefont
  {Girod}}, \bibinfo {author} {\bibfnamefont {C.}~\bibnamefont {Lorc{\'e}}},
  \bibinfo {author} {\bibfnamefont {P.}~\bibnamefont {Schweitzer}}, \ and\
  \bibinfo {author} {\bibfnamefont {P.~E.}\ \bibnamefont {Shanahan}},\ }\href
  {\doibase 10.1103/RevModPhys.95.041002} {\bibfield  {journal} {\bibinfo
  {journal} {Rev. Mod. Phys.}\ }\textbf {\bibinfo {volume} {95}},\ \bibinfo
  {pages} {041002} (\bibinfo {year} {2023})},\ \Eprint
  {http://arxiv.org/abs/2303.08347} {arXiv:2303.08347 [hep-ph]} \BibitemShut
  {NoStop}%
\bibitem [{\citenamefont {Lorc{\'e}}\ and\ \citenamefont
  {Schweitzer}(2025)}]{Lorce:2025oot}%
  \BibitemOpen
  \bibfield  {author} {\bibinfo {author} {\bibfnamefont {C.}~\bibnamefont
  {Lorc{\'e}}}\ and\ \bibinfo {author} {\bibfnamefont {P.}~\bibnamefont
  {Schweitzer}},\ }\href {\doibase 10.5506/APhysPolB.56.3-A17} {\bibfield
  {journal} {\bibinfo  {journal} {Acta Phys. Polon. B}\ }\textbf {\bibinfo
  {volume} {56}},\ \bibinfo {pages} {3} (\bibinfo {year} {2025})},\ \Eprint
  {http://arxiv.org/abs/2501.04622} {arXiv:2501.04622 [hep-ph]} \BibitemShut
  {NoStop}%
\bibitem [{\citenamefont {Polyakov}\ and\ \citenamefont
  {Sun}(2019)}]{Polyakov:2019lbq}%
  \BibitemOpen
  \bibfield  {author} {\bibinfo {author} {\bibfnamefont {M.~V.}\ \bibnamefont
  {Polyakov}}\ and\ \bibinfo {author} {\bibfnamefont {B.-D.}\ \bibnamefont
  {Sun}},\ }\href {\doibase 10.1103/PhysRevD.100.036003} {\bibfield  {journal}
  {\bibinfo  {journal} {Phys. Rev. D}\ }\textbf {\bibinfo {volume} {100}},\
  \bibinfo {pages} {036003} (\bibinfo {year} {2019})},\ \Eprint
  {http://arxiv.org/abs/1903.02738} {arXiv:1903.02738 [hep-ph]} \BibitemShut
  {NoStop}%
\bibitem [{\citenamefont {Kim}\ and\ \citenamefont {Sun}(2021)}]{Kim:2020lrs}%
  \BibitemOpen
  \bibfield  {author} {\bibinfo {author} {\bibfnamefont {J.-Y.}\ \bibnamefont
  {Kim}}\ and\ \bibinfo {author} {\bibfnamefont {B.-D.}\ \bibnamefont {Sun}},\
  }\href {\doibase 10.1140/epjc/s10052-021-08852-z} {\bibfield  {journal}
  {\bibinfo  {journal} {Eur. Phys. J. C}\ }\textbf {\bibinfo {volume} {81}},\
  \bibinfo {pages} {85} (\bibinfo {year} {2021})},\ \Eprint
  {http://arxiv.org/abs/2011.00292} {arXiv:2011.00292 [hep-ph]} \BibitemShut
  {NoStop}%
\bibitem [{\citenamefont {Kim}\ \emph {et~al.}(2023{\natexlab{a}})\citenamefont
  {Kim}, \citenamefont {Sun}, \citenamefont {Fu},\ and\ \citenamefont
  {Kim}}]{Kim:2022wkc}%
  \BibitemOpen
  \bibfield  {author} {\bibinfo {author} {\bibfnamefont {J.-Y.}\ \bibnamefont
  {Kim}}, \bibinfo {author} {\bibfnamefont {B.-D.}\ \bibnamefont {Sun}},
  \bibinfo {author} {\bibfnamefont {D.}~\bibnamefont {Fu}}, \ and\ \bibinfo
  {author} {\bibfnamefont {H.-C.}\ \bibnamefont {Kim}},\ }\href {\doibase
  10.1103/PhysRevD.107.054007} {\bibfield  {journal} {\bibinfo  {journal}
  {Phys. Rev. D}\ }\textbf {\bibinfo {volume} {107}},\ \bibinfo {pages}
  {054007} (\bibinfo {year} {2023}{\natexlab{a}})},\ \Eprint
  {http://arxiv.org/abs/2208.01240} {arXiv:2208.01240 [hep-ph]} \BibitemShut
  {NoStop}%
\bibitem [{\citenamefont {Freese}\ and\ \citenamefont
  {Cosyn}(2022)}]{Freese:2022yur}%
  \BibitemOpen
  \bibfield  {author} {\bibinfo {author} {\bibfnamefont {A.}~\bibnamefont
  {Freese}}\ and\ \bibinfo {author} {\bibfnamefont {W.}~\bibnamefont {Cosyn}},\
  }\href {\doibase 10.1103/PhysRevD.106.114013} {\bibfield  {journal} {\bibinfo
   {journal} {Phys. Rev. D}\ }\textbf {\bibinfo {volume} {106}},\ \bibinfo
  {pages} {114013} (\bibinfo {year} {2022})},\ \Eprint
  {http://arxiv.org/abs/2207.10787} {arXiv:2207.10787 [hep-ph]} \BibitemShut
  {NoStop}%
\bibitem [{\citenamefont {Dehghan}\ \emph {et~al.}(2023)\citenamefont
  {Dehghan}, \citenamefont {Azizi},\ and\ \citenamefont
  {{\"O}zdem}}]{Dehghan:2023ytx}%
  \BibitemOpen
  \bibfield  {author} {\bibinfo {author} {\bibfnamefont {Z.}~\bibnamefont
  {Dehghan}}, \bibinfo {author} {\bibfnamefont {K.}~\bibnamefont {Azizi}}, \
  and\ \bibinfo {author} {\bibfnamefont {U.}~\bibnamefont {{\"O}zdem}},\ }\href
  {\doibase 10.1103/PhysRevD.108.094037} {\bibfield  {journal} {\bibinfo
  {journal} {Phys. Rev. D}\ }\textbf {\bibinfo {volume} {108}},\ \bibinfo
  {pages} {094037} (\bibinfo {year} {2023})},\ \Eprint
  {http://arxiv.org/abs/2307.14880} {arXiv:2307.14880 [hep-ph]} \BibitemShut
  {NoStop}%
\bibitem [{\citenamefont {Kim}(2022{\natexlab{a}})}]{Kim:2022bwn}%
  \BibitemOpen
  \bibfield  {author} {\bibinfo {author} {\bibfnamefont {J.-Y.}\ \bibnamefont
  {Kim}},\ }\href {\doibase 10.1016/j.physletb.2022.137442} {\bibfield
  {journal} {\bibinfo  {journal} {Phys. Lett. B}\ }\textbf {\bibinfo {volume}
  {834}},\ \bibinfo {pages} {137442} (\bibinfo {year} {2022}{\natexlab{a}})},\
  \Eprint {http://arxiv.org/abs/2206.10202} {arXiv:2206.10202 [hep-ph]}
  \BibitemShut {NoStop}%
\bibitem [{\citenamefont {Kim}\ \emph {et~al.}(2023{\natexlab{b}})\citenamefont
  {Kim}, \citenamefont {Won}, \citenamefont {Goity},\ and\ \citenamefont
  {Weiss}}]{Kim:2023xvw}%
  \BibitemOpen
  \bibfield  {author} {\bibinfo {author} {\bibfnamefont {J.-Y.}\ \bibnamefont
  {Kim}}, \bibinfo {author} {\bibfnamefont {H.-Y.}\ \bibnamefont {Won}},
  \bibinfo {author} {\bibfnamefont {J.~L.}\ \bibnamefont {Goity}}, \ and\
  \bibinfo {author} {\bibfnamefont {C.}~\bibnamefont {Weiss}},\ }\href@noop {}
  {\  (\bibinfo {year} {2023}{\natexlab{b}})},\ \Eprint
  {http://arxiv.org/abs/2304.08575} {arXiv:2304.08575 [hep-ph]} \BibitemShut
  {NoStop}%
\bibitem [{\citenamefont {Kim}(2023)}]{Kim:2023yhp}%
  \BibitemOpen
  \bibfield  {author} {\bibinfo {author} {\bibfnamefont {J.-Y.}\ \bibnamefont
  {Kim}},\ }\href {\doibase 10.1103/PhysRevD.108.034024} {\bibfield  {journal}
  {\bibinfo  {journal} {Phys. Rev. D}\ }\textbf {\bibinfo {volume} {108}},\
  \bibinfo {pages} {034024} (\bibinfo {year} {2023})},\ \Eprint
  {http://arxiv.org/abs/2305.12714} {arXiv:2305.12714 [hep-ph]} \BibitemShut
  {NoStop}%
\bibitem [{\citenamefont {Alharazin}\ \emph {et~al.}(2024)\citenamefont
  {Alharazin}, \citenamefont {Sun}, \citenamefont {Epelbaum}, \citenamefont
  {Gegelia},\ and\ \citenamefont {Mei{\ss}ner}}]{Alharazin:2023zzc}%
  \BibitemOpen
  \bibfield  {author} {\bibinfo {author} {\bibfnamefont {H.}~\bibnamefont
  {Alharazin}}, \bibinfo {author} {\bibfnamefont {B.~D.}\ \bibnamefont {Sun}},
  \bibinfo {author} {\bibfnamefont {E.}~\bibnamefont {Epelbaum}}, \bibinfo
  {author} {\bibfnamefont {J.}~\bibnamefont {Gegelia}}, \ and\ \bibinfo
  {author} {\bibfnamefont {U.~G.}\ \bibnamefont {Mei{\ss}ner}},\ }\href
  {\doibase 10.1007/JHEP03(2024)007} {\bibfield  {journal} {\bibinfo  {journal}
  {JHEP}\ }\textbf {\bibinfo {volume} {03}},\ \bibinfo {pages} {007} (\bibinfo
  {year} {2024})},\ \Eprint {http://arxiv.org/abs/2312.05193} {arXiv:2312.05193
  [hep-ph]} \BibitemShut {NoStop}%
\bibitem [{\citenamefont {Kim}\ and\ \citenamefont
  {Weiss}(2025)}]{Kim:2025ilc}%
  \BibitemOpen
  \bibfield  {author} {\bibinfo {author} {\bibfnamefont {J.-Y.}\ \bibnamefont
  {Kim}}\ and\ \bibinfo {author} {\bibfnamefont {C.}~\bibnamefont {Weiss}},\
  }\href@noop {} {\  (\bibinfo {year} {2025})},\ \Eprint
  {http://arxiv.org/abs/2507.18402} {arXiv:2507.18402 [hep-ph]} \BibitemShut
  {NoStop}%
\bibitem [{\citenamefont {Kim}(2025)}]{Kim:2025ves}%
  \BibitemOpen
  \bibfield  {author} {\bibinfo {author} {\bibfnamefont {J.-Y.}\ \bibnamefont
  {Kim}},\ }\href@noop {} {\  (\bibinfo {year} {2025})},\ \Eprint
  {http://arxiv.org/abs/2508.11491} {arXiv:2508.11491 [hep-ph]} \BibitemShut
  {NoStop}%
\bibitem [{\citenamefont {Polyakov}\ and\ \citenamefont
  {Son}(2018)}]{Polyakov:2018exb}%
  \BibitemOpen
  \bibfield  {author} {\bibinfo {author} {\bibfnamefont {M.~V.}\ \bibnamefont
  {Polyakov}}\ and\ \bibinfo {author} {\bibfnamefont {H.-D.}\ \bibnamefont
  {Son}},\ }\href {\doibase 10.1007/JHEP09(2018)156} {\bibfield  {journal}
  {\bibinfo  {journal} {JHEP}\ }\textbf {\bibinfo {volume} {09}},\ \bibinfo
  {pages} {156} (\bibinfo {year} {2018})},\ \Eprint
  {http://arxiv.org/abs/1808.00155} {arXiv:1808.00155 [hep-ph]} \BibitemShut
  {NoStop}%
\bibitem [{\citenamefont {Won}\ \emph {et~al.}(2024{\natexlab{a}})\citenamefont
  {Won}, \citenamefont {Kim},\ and\ \citenamefont {Kim}}]{Won:2023cyd}%
  \BibitemOpen
  \bibfield  {author} {\bibinfo {author} {\bibfnamefont {H.-Y.}\ \bibnamefont
  {Won}}, \bibinfo {author} {\bibfnamefont {H.-C.}\ \bibnamefont {Kim}}, \ and\
  \bibinfo {author} {\bibfnamefont {J.-Y.}\ \bibnamefont {Kim}},\ }\href
  {\doibase 10.1016/j.physletb.2024.138489} {\bibfield  {journal} {\bibinfo
  {journal} {Phys. Lett. B}\ }\textbf {\bibinfo {volume} {850}},\ \bibinfo
  {pages} {138489} (\bibinfo {year} {2024}{\natexlab{a}})},\ \Eprint
  {http://arxiv.org/abs/2302.02974} {arXiv:2302.02974 [hep-ph]} \BibitemShut
  {NoStop}%
\bibitem [{\citenamefont {Won}\ \emph {et~al.}(2024{\natexlab{b}})\citenamefont
  {Won}, \citenamefont {Kim},\ and\ \citenamefont {Kim}}]{Won:2023zmf}%
  \BibitemOpen
  \bibfield  {author} {\bibinfo {author} {\bibfnamefont {H.-Y.}\ \bibnamefont
  {Won}}, \bibinfo {author} {\bibfnamefont {H.-C.}\ \bibnamefont {Kim}}, \ and\
  \bibinfo {author} {\bibfnamefont {J.-Y.}\ \bibnamefont {Kim}},\ }\href
  {\doibase 10.1007/JHEP05(2024)173} {\bibfield  {journal} {\bibinfo  {journal}
  {JHEP}\ }\textbf {\bibinfo {volume} {05}},\ \bibinfo {pages} {173} (\bibinfo
  {year} {2024}{\natexlab{b}})},\ \Eprint {http://arxiv.org/abs/2310.04670}
  {arXiv:2310.04670 [hep-ph]} \BibitemShut {NoStop}%
\bibitem [{\citenamefont {Freese}(2025)}]{Freese:2024rkr}%
  \BibitemOpen
  \bibfield  {author} {\bibinfo {author} {\bibfnamefont {A.}~\bibnamefont
  {Freese}},\ }\href {\doibase 10.1103/PhysRevD.111.034047} {\bibfield
  {journal} {\bibinfo  {journal} {Phys. Rev. D}\ }\textbf {\bibinfo {volume}
  {111}},\ \bibinfo {pages} {034047} (\bibinfo {year} {2025})},\ \Eprint
  {http://arxiv.org/abs/2412.09664} {arXiv:2412.09664 [hep-ph]} \BibitemShut
  {NoStop}%
\bibitem [{\citenamefont {Ji}\ and\ \citenamefont
  {Yang}(2025{\natexlab{a}})}]{Ji:2025gsq}%
  \BibitemOpen
  \bibfield  {author} {\bibinfo {author} {\bibfnamefont {X.}~\bibnamefont
  {Ji}}\ and\ \bibinfo {author} {\bibfnamefont {C.}~\bibnamefont {Yang}},\
  }\href@noop {} {\  (\bibinfo {year} {2025}{\natexlab{a}})},\ \Eprint
  {http://arxiv.org/abs/2503.01991} {arXiv:2503.01991 [hep-ph]} \BibitemShut
  {NoStop}%
\bibitem [{\citenamefont {Lorc{\'e}}(2020)}]{Lorce:2020onh}%
  \BibitemOpen
  \bibfield  {author} {\bibinfo {author} {\bibfnamefont {C.}~\bibnamefont
  {Lorc{\'e}}},\ }\href {\doibase 10.1103/PhysRevLett.125.232002} {\bibfield
  {journal} {\bibinfo  {journal} {Phys. Rev. Lett.}\ }\textbf {\bibinfo
  {volume} {125}},\ \bibinfo {pages} {232002} (\bibinfo {year} {2020})},\
  \Eprint {http://arxiv.org/abs/2007.05318} {arXiv:2007.05318 [hep-ph]}
  \BibitemShut {NoStop}%
\bibitem [{\citenamefont {Won}\ and\ \citenamefont
  {Lorc{\'e}}(2025)}]{Won:2025dgc}%
  \BibitemOpen
  \bibfield  {author} {\bibinfo {author} {\bibfnamefont {H.-Y.}\ \bibnamefont
  {Won}}\ and\ \bibinfo {author} {\bibfnamefont {C.}~\bibnamefont
  {Lorc{\'e}}},\ }\href {\doibase 10.1103/PhysRevD.111.094021} {\bibfield
  {journal} {\bibinfo  {journal} {Phys. Rev. D}\ }\textbf {\bibinfo {volume}
  {111}},\ \bibinfo {pages} {094021} (\bibinfo {year} {2025})},\ \Eprint
  {http://arxiv.org/abs/2503.07382} {arXiv:2503.07382 [hep-ph]} \BibitemShut
  {NoStop}%
\bibitem [{\citenamefont {Lorc{\'e}}\ \emph {et~al.}(2022)\citenamefont
  {Lorc{\'e}}, \citenamefont {Schweitzer},\ and\ \citenamefont
  {Tezgin}}]{Lorce:2022cle}%
  \BibitemOpen
  \bibfield  {author} {\bibinfo {author} {\bibfnamefont {C.}~\bibnamefont
  {Lorc{\'e}}}, \bibinfo {author} {\bibfnamefont {P.}~\bibnamefont
  {Schweitzer}}, \ and\ \bibinfo {author} {\bibfnamefont {K.}~\bibnamefont
  {Tezgin}},\ }\href {\doibase 10.1103/PhysRevD.106.014012} {\bibfield
  {journal} {\bibinfo  {journal} {Phys. Rev. D}\ }\textbf {\bibinfo {volume}
  {106}},\ \bibinfo {pages} {014012} (\bibinfo {year} {2022})},\ \Eprint
  {http://arxiv.org/abs/2202.01192} {arXiv:2202.01192 [hep-ph]} \BibitemShut
  {NoStop}%
\bibitem [{\citenamefont {Kim}\ and\ \citenamefont
  {Kim}(2021{\natexlab{b}})}]{Kim:2021kum}%
  \BibitemOpen
  \bibfield  {author} {\bibinfo {author} {\bibfnamefont {J.-Y.}\ \bibnamefont
  {Kim}}\ and\ \bibinfo {author} {\bibfnamefont {H.-C.}\ \bibnamefont {Kim}},\
  }\href {\doibase 10.1103/PhysRevD.104.074003} {\bibfield  {journal} {\bibinfo
   {journal} {Phys. Rev. D}\ }\textbf {\bibinfo {volume} {104}},\ \bibinfo
  {pages} {074003} (\bibinfo {year} {2021}{\natexlab{b}})},\ \Eprint
  {http://arxiv.org/abs/2106.10986} {arXiv:2106.10986 [hep-ph]} \BibitemShut
  {NoStop}%
\bibitem [{\citenamefont {Kim}(2022{\natexlab{b}})}]{Kim:2022bia}%
  \BibitemOpen
  \bibfield  {author} {\bibinfo {author} {\bibfnamefont {J.-Y.}\ \bibnamefont
  {Kim}},\ }\href {\doibase 10.1103/PhysRevD.106.014022} {\bibfield  {journal}
  {\bibinfo  {journal} {Phys. Rev. D}\ }\textbf {\bibinfo {volume} {106}},\
  \bibinfo {pages} {014022} (\bibinfo {year} {2022}{\natexlab{b}})},\ \Eprint
  {http://arxiv.org/abs/2204.08248} {arXiv:2204.08248 [hep-ph]} \BibitemShut
  {NoStop}%
\bibitem [{\citenamefont {Varshalovich}\ \emph {et~al.}(1988)\citenamefont
  {Varshalovich}, \citenamefont {Moskalev},\ and\ \citenamefont
  {Khersonskii}}]{Varshalovich:1988ifq}%
  \BibitemOpen
  \bibfield  {author} {\bibinfo {author} {\bibfnamefont {D.~A.}\ \bibnamefont
  {Varshalovich}}, \bibinfo {author} {\bibfnamefont {A.~N.}\ \bibnamefont
  {Moskalev}}, \ and\ \bibinfo {author} {\bibfnamefont {V.~K.}\ \bibnamefont
  {Khersonskii}},\ }\href {\doibase 10.1142/0270} {\emph {\bibinfo {title}
  {{Quantum Theory of Angular Momentum}: {Irreducible Tensors, Spherical
  Harmonics, Vector Coupling Coefficients, 3nj Symbols}}}}\ (\bibinfo
  {publisher} {World Scientific Publishing Company},\ \bibinfo {year}
  {1988})\BibitemShut {NoStop}%
\bibitem [{\citenamefont {Cotogno}\ \emph {et~al.}(2020)\citenamefont
  {Cotogno}, \citenamefont {Lorc{\'e}}, \citenamefont {Lowdon},\ and\
  \citenamefont {Morales}}]{Cotogno:2019vjb}%
  \BibitemOpen
  \bibfield  {author} {\bibinfo {author} {\bibfnamefont {S.}~\bibnamefont
  {Cotogno}}, \bibinfo {author} {\bibfnamefont {C.}~\bibnamefont {Lorc{\'e}}},
  \bibinfo {author} {\bibfnamefont {P.}~\bibnamefont {Lowdon}}, \ and\ \bibinfo
  {author} {\bibfnamefont {M.}~\bibnamefont {Morales}},\ }\href {\doibase
  10.1103/PhysRevD.101.056016} {\bibfield  {journal} {\bibinfo  {journal}
  {Phys. Rev. D}\ }\textbf {\bibinfo {volume} {101}},\ \bibinfo {pages}
  {056016} (\bibinfo {year} {2020})},\ \Eprint
  {http://arxiv.org/abs/1912.08749} {arXiv:1912.08749 [hep-ph]} \BibitemShut
  {NoStop}%
\bibitem [{\citenamefont {Panteleeva}\ and\ \citenamefont
  {Polyakov}(2020)}]{Panteleeva:2020ejw}%
  \BibitemOpen
  \bibfield  {author} {\bibinfo {author} {\bibfnamefont {J.~Y.}\ \bibnamefont
  {Panteleeva}}\ and\ \bibinfo {author} {\bibfnamefont {M.~V.}\ \bibnamefont
  {Polyakov}},\ }\href {\doibase 10.1016/j.physletb.2020.135707} {\bibfield
  {journal} {\bibinfo  {journal} {Phys. Lett. B}\ }\textbf {\bibinfo {volume}
  {809}},\ \bibinfo {pages} {135707} (\bibinfo {year} {2020})},\ \Eprint
  {http://arxiv.org/abs/2004.02912} {arXiv:2004.02912 [hep-ph]} \BibitemShut
  {NoStop}%
\bibitem [{\citenamefont {Landau}\ \emph {et~al.}(1986)\citenamefont {Landau},
  \citenamefont {Landau}, \citenamefont {Lifshits}, \citenamefont {Kosevich},
  \citenamefont {Lifshitz},\ and\ \citenamefont
  {Pitaevskii}}]{landau1986theory}%
  \BibitemOpen
  \bibfield  {author} {\bibinfo {author} {\bibfnamefont {L.}~\bibnamefont
  {Landau}}, \bibinfo {author} {\bibfnamefont {L.}~\bibnamefont {Landau}},
  \bibinfo {author} {\bibfnamefont {E.}~\bibnamefont {Lifshits}}, \bibinfo
  {author} {\bibfnamefont {A.}~\bibnamefont {Kosevich}}, \bibinfo {author}
  {\bibfnamefont {E.}~\bibnamefont {Lifshitz}}, \ and\ \bibinfo {author}
  {\bibfnamefont {L.}~\bibnamefont {Pitaevskii}},\ }\href
  {https://books.google.com/books?id=tpY-VkwCkAIC} {\emph {\bibinfo {title}
  {Theory of Elasticity: Volume 7}}},\ Course of theoretical physics\ (\bibinfo
   {publisher} {Butterworth-Heinemann},\ \bibinfo {year} {1986})\BibitemShut
  {NoStop}%
\bibitem [{\citenamefont {Ji}(1995)}]{Ji:1995sv}%
  \BibitemOpen
  \bibfield  {author} {\bibinfo {author} {\bibfnamefont {X.-D.}\ \bibnamefont
  {Ji}},\ }\href {\doibase 10.1103/PhysRevD.52.271} {\bibfield  {journal}
  {\bibinfo  {journal} {Phys. Rev. D}\ }\textbf {\bibinfo {volume} {52}},\
  \bibinfo {pages} {271} (\bibinfo {year} {1995})},\ \Eprint
  {http://arxiv.org/abs/hep-ph/9502213} {arXiv:hep-ph/9502213} \BibitemShut
  {NoStop}%
\bibitem [{\citenamefont {Liu}(2021)}]{Liu:2021gco}%
  \BibitemOpen
  \bibfield  {author} {\bibinfo {author} {\bibfnamefont {K.-F.}\ \bibnamefont
  {Liu}},\ }\href {\doibase 10.1103/PhysRevD.104.076010} {\bibfield  {journal}
  {\bibinfo  {journal} {Phys. Rev. D}\ }\textbf {\bibinfo {volume} {104}},\
  \bibinfo {pages} {076010} (\bibinfo {year} {2021})},\ \Eprint
  {http://arxiv.org/abs/2103.15768} {arXiv:2103.15768 [hep-ph]} \BibitemShut
  {NoStop}%
\bibitem [{\citenamefont {Lorc{\'e}}(2018)}]{Lorce:2017xzd}%
  \BibitemOpen
  \bibfield  {author} {\bibinfo {author} {\bibfnamefont {C.}~\bibnamefont
  {Lorc{\'e}}},\ }\href {\doibase 10.1140/epjc/s10052-018-5561-2} {\bibfield
  {journal} {\bibinfo  {journal} {Eur. Phys. J. C}\ }\textbf {\bibinfo {volume}
  {78}},\ \bibinfo {pages} {120} (\bibinfo {year} {2018})},\ \Eprint
  {http://arxiv.org/abs/1706.05853} {arXiv:1706.05853 [hep-ph]} \BibitemShut
  {NoStop}%
\bibitem [{\citenamefont {Hatta}\ \emph {et~al.}(2018)\citenamefont {Hatta},
  \citenamefont {Rajan},\ and\ \citenamefont {Tanaka}}]{Hatta:2018sqd}%
  \BibitemOpen
  \bibfield  {author} {\bibinfo {author} {\bibfnamefont {Y.}~\bibnamefont
  {Hatta}}, \bibinfo {author} {\bibfnamefont {A.}~\bibnamefont {Rajan}}, \ and\
  \bibinfo {author} {\bibfnamefont {K.}~\bibnamefont {Tanaka}},\ }\href
  {\doibase 10.1007/JHEP12(2018)008} {\bibfield  {journal} {\bibinfo  {journal}
  {JHEP}\ }\textbf {\bibinfo {volume} {12}},\ \bibinfo {pages} {008} (\bibinfo
  {year} {2018})},\ \Eprint {http://arxiv.org/abs/1810.05116} {arXiv:1810.05116
  [hep-ph]} \BibitemShut {NoStop}%
\bibitem [{\citenamefont {Ji}\ and\ \citenamefont
  {Yang}(2025{\natexlab{b}})}]{Ji:2025qax}%
  \BibitemOpen
  \bibfield  {author} {\bibinfo {author} {\bibfnamefont {X.}~\bibnamefont
  {Ji}}\ and\ \bibinfo {author} {\bibfnamefont {C.}~\bibnamefont {Yang}},\
  }\href@noop {} {\  (\bibinfo {year} {2025}{\natexlab{b}})},\ \Eprint
  {http://arxiv.org/abs/2508.16727} {arXiv:2508.16727 [hep-ph]} \BibitemShut
  {NoStop}%
\bibitem [{\citenamefont {Metz}\ \emph {et~al.}(2020)\citenamefont {Metz},
  \citenamefont {Pasquini},\ and\ \citenamefont {Rodini}}]{Metz:2020vxd}%
  \BibitemOpen
  \bibfield  {author} {\bibinfo {author} {\bibfnamefont {A.}~\bibnamefont
  {Metz}}, \bibinfo {author} {\bibfnamefont {B.}~\bibnamefont {Pasquini}}, \
  and\ \bibinfo {author} {\bibfnamefont {S.}~\bibnamefont {Rodini}},\ }\href
  {\doibase 10.1103/PhysRevD.102.114042} {\bibfield  {journal} {\bibinfo
  {journal} {Phys. Rev. D}\ }\textbf {\bibinfo {volume} {102}},\ \bibinfo
  {pages} {114042} (\bibinfo {year} {2020})},\ \Eprint
  {http://arxiv.org/abs/2006.11171} {arXiv:2006.11171 [hep-ph]} \BibitemShut
  {NoStop}%
\bibitem [{\citenamefont {Tanaka}(2018)}]{Tanaka:2018wea}%
  \BibitemOpen
  \bibfield  {author} {\bibinfo {author} {\bibfnamefont {K.}~\bibnamefont
  {Tanaka}},\ }\href {\doibase 10.1103/PhysRevD.98.034009} {\bibfield
  {journal} {\bibinfo  {journal} {Phys. Rev. D}\ }\textbf {\bibinfo {volume}
  {98}},\ \bibinfo {pages} {034009} (\bibinfo {year} {2018})},\ \Eprint
  {http://arxiv.org/abs/1806.10591} {arXiv:1806.10591 [hep-ph]} \BibitemShut
  {NoStop}%
\bibitem [{\citenamefont {Cosyn}\ \emph {et~al.}(2019)\citenamefont {Cosyn},
  \citenamefont {Cotogno}, \citenamefont {Freese},\ and\ \citenamefont
  {Lorc{\'e}}}]{Cosyn:2019aio}%
  \BibitemOpen
  \bibfield  {author} {\bibinfo {author} {\bibfnamefont {W.}~\bibnamefont
  {Cosyn}}, \bibinfo {author} {\bibfnamefont {S.}~\bibnamefont {Cotogno}},
  \bibinfo {author} {\bibfnamefont {A.}~\bibnamefont {Freese}}, \ and\ \bibinfo
  {author} {\bibfnamefont {C.}~\bibnamefont {Lorc{\'e}}},\ }\href {\doibase
  10.1140/epjc/s10052-019-6981-3} {\bibfield  {journal} {\bibinfo  {journal}
  {Eur. Phys. J. C}\ }\textbf {\bibinfo {volume} {79}},\ \bibinfo {pages} {476}
  (\bibinfo {year} {2019})},\ \Eprint {http://arxiv.org/abs/1903.00408}
  {arXiv:1903.00408 [hep-ph]} \BibitemShut {NoStop}%
\bibitem [{\citenamefont {Carlson}\ and\ \citenamefont
  {Vanderhaeghen}(2009)}]{Carlson:2009ovh}%
  \BibitemOpen
  \bibfield  {author} {\bibinfo {author} {\bibfnamefont {C.~E.}\ \bibnamefont
  {Carlson}}\ and\ \bibinfo {author} {\bibfnamefont {M.}~\bibnamefont
  {Vanderhaeghen}},\ }\href {\doibase 10.1140/epja/i2009-10800-0} {\bibfield
  {journal} {\bibinfo  {journal} {Eur. Phys. J. A}\ }\textbf {\bibinfo {volume}
  {41}},\ \bibinfo {pages} {1} (\bibinfo {year} {2009})},\ \Eprint
  {http://arxiv.org/abs/0807.4537} {arXiv:0807.4537 [hep-ph]} \BibitemShut
  {NoStop}%
\end{thebibliography}%

\end{document}